# Room temperature spin-orbit torque efficiency in sputtered low-temperature superconductor $\delta$-TaN

Przemyslaw Wojciech Swatek[1§*], Xudong Hang[1§], Yihong Fan[1§], Wei Jiang[1§], Hwanhui Yun[2§], Deyuan Lyu[1], Delin Zhang[1], Thomas J Peterson[3], Protyush Sahu[3], Onri Jay Benally[1], Zach Cresswell[2], Jinming Liu[1], Rabindra Pahari[1], Daniel Kukla[3], Tony Low[1], K. Andre Mkhoyan[2], Jian-Ping Wang[1*]

[1] *Department of Electrical and Computer Engineering, University of Minnesota,*

*Minneapolis, Minnesota 55455, USA*

[2] *Department of Chemical Engineering and Materials Science, University of Minnesota, Minneapolis, Minnesota 55455, USA*

[3]*School of Physics and Astronomy, University of Minnesota, 116 Church Street SE, Minneapolis, MN 55455, USA*

*Corresponding authors: jpwang@umn.edu (J.P.W.) and pwswatek@gmail.com (P.W.S.)*
*§ Equal contribution authors*

In the course of searching for new promising topological materials for applications in future topological electronics, we evaluated spin-orbit torques (SOT) in high quality sputtered $\delta$-TaN/Co$_{20}$Fe$_{60}$B$_{20}$ devices through spin-torque ferromagnetic resonance (ST-FMR) and spin pumping measurements. From the ST-FMR characterization we observed a significant linewidth modulation in the magnetic Co$_{20}$Fe$_{60}$B$_{20}$ layer attributed to the charge to spin conversion generated from the $\delta$-TaN layer. Remarkably, the spin torque efficiency determined from ST-FMR and spin pumping measurements is as large as $\Theta = 0.034$ and $0.031$, respectively. These values are over two times larger than for $\alpha$-Ta, but almost five times lower than for β-Ta, which can be attributed to the low room temperature (RT) electrical resistivity ~74 $\mu\Omega$-cm in $\delta$-TaN. A large spin diffusion length of at least ~8 nm is estimated, which is comparable to the spin diffusion length in pure Ta. Comprehensive experimental analysis, together with density functional theory (DFT) calculations, indicates that the origin of the pronounced SOT effect in $\delta$-TaN **can be mostly related to a significant contribution from the Berry curvature associated with the presence of a topically nontrivial electronic band structure in the vicinity of the Fermi level ($E_F$). Through additional detailed theoretical analysis, we also found that an isostructural allotrope of superconducting $\delta$-TaN phase, the simple hexagonal structure $\theta$-TaN, has larger Berry curvature, and that, together with expected lower conductivity, it can also be a promising candidate for exploring a new generation of Spin-Orbit Torque Magnetic Random Access Memory (SOT-MRAM) as a cheap, temperature stable, and highly efficient spin currents sources.**

## I. INTRODUCTION

The discovery of various exotic topological states that can be experimentally realized in semimetals has ignited intensive studies. Besides their unprecedented importance for fundamental science, they offer intriguing possibilities for device design with revolutionizing low-power computation capabilities, as well as laser technology [1-9]. Non-magnetic topological semimetals (TSs) and insulators (TIs) are at the top among promising materials for spintronic applications, especially in the context of the new generation of highly efficient SOT-MRAM devices [10-21]. Recent theories focusing on a variety of possible symmetries in condensed-matter physics have expanded the zoo of known topological quasiparticle excitations [22]. In the context of highly efficient spin-orbit torque (SOT) materials, degeneracies of energy bands including three-, four-, and six-fold chiral fermions deserve special attention, since degeneracies near the Fermi level ($E_F$) can lead to greatly enhanced Berry curvature, which governs the intrinsic spin Hall conductivity (SHC) [13,22-30]. However, despite large charge-to-spin and spin-to-charge conversion efficiencies of most well-known TSs and TIs, they usually have large resistivity ($10^3$-$10^5$ $\mu\Omega$-cm) compared to commonly studied SOT generators based on heavy metals (10-300 $\mu\Omega$-cm). The key challenge is to find new materials with low resistivity that provide a balanced combination of both *i*) topologically non-trivial and trivial electronic states near $E_F$ and *ii*) tunable electronic structure to obtain a large and efficient charge-to-spin conversion figure of merit.

Pure tantalum has been adopted in many SOT experiments because of its relatively large spin Hall angle (SHA) [10,31]. Two phases of solid Ta can exist in two different crystal structures: α-Ta, which is body-centered-cubic, and metastable-tetragonal β-Ta [32]. Due to different crystal symmetries and related electronic band structures, the transport properties of each phase are quite different. The SHC of β-Ta is −389 ($\hbar/e$) S/cm, while that of α-Ta is −142 ($\hbar/e$) S/cm [32]. Based on experimental results, the resistivity of *β*-Ta is around 150 - 200 $\mu\Omega$-cm, which is approximately four to six times as large as that of *α*-Ta. Therefore, the SHA of α-Ta is estimated to be around −0.014, while for *β*-Ta it is ~ −0.16, both of which are supported by many experimental investigations [33-43]. The crucial question here is how can we increase SHC and/or lower the charge intrinsic resistivity to maximize the value of SHA to obtain a new material of technological interest for spintronic applications. One approach is to try improving the electronic structure of pure *α*-Ta to create alternative non-trivial, highly spin-polarized electronic states that will occur in the presence of trivial bulk states [11,12].

Here we demonstrate that by incorporating nitrogen into tantalum, a pronounced SOT effect compared to pure *α*-Ta, with a relatively low resistivity, are attained in *δ*-TaN phase [23,44-55]. The room temperature value of the SHA of 0.034 was determined using ST-FMR and spin-pumping methods, highlighting a good spin-to-charge conversion efficiency. From the comparison between spin pumping and ST-FMR measurements, we derive a large spin diffusion length of ~ 8 nm in δ-TaN. Our theoretical calculations suggest that the experimental SOT findings are largely associated with enhanced Berry curvature due to non-trivial electronic structures along some high-symmetry *k*-paths in the Brillouin zone (BZ). Importantly, due to the crystal symmetry and moderate spin-orbit coupling (SOC), around 50% of the total SHC generated in δ-TaN layer can be attributed to the $\sigma_{zy}^{z}$ spin out-of-plane polarization component, as suggested by our DFT results. Therefore, our experimental and theoretical results provide valuable insight into the spin transport in δ-TaN and open the door toward further engineering efficient and reliable SOT-MRAM devices based on Ta.

## II. METHODS

All samples with structure MgO/TaN(10) (thickness in nanometers) were deposited on the MgO single crystal substrates with the temperature at 400 ºC and the base pressure <3.5×10$^{-8}$ Torr using a facing-target sputtering (FTS) system, which produces high-quality thin-film samples that are free of radiation damage. The working principle of the system can be found in previous reports [56]. The stacks of $Co_{20}Fe_{60}B_{20}$ (hereafter referred to CFB) (2.5-6)/MgO(2)/Ta(2) were grown at room temperature on such obtained MgO(001)/TaN(10) by a six-target Shamrock magnetron sputtering system under a base pressure less than <5×10$^{-8}$ Torr. X-ray diffraction (XRD) and reflectometry (XRR) experiments were performed on a Rigaku Smartlab XE high-resolution diffractometer with Cu K-alpha1 radiation (wavelength 1.5406 Å). For the fitting purpose, the TaN layer thickness in the XRR samples is increased to more than 20 nm to obtain multiple peaks. Then, the samples were characterized in a PHI 5000 Versaprobe III photoelectron spectrometer (XPS) with a monochromatic *Al Kα* x-ray with the energy of 1486.6 eV. The pass energy of 280 eV and 55 eV were used to collect survey spectra and core-level spectra, respectively. Cross-sectional transmission electron microscopy (TEM) samples were prepared by using a focused-ion beam (FIB) lift-out method using an FEI Helios Nanolab G4 dual-beam FIB. Scanning transmission electron microscopy (STEM) imaging and spectroscopy experiments were carried out using an aberration-corrected FEI Titan G2 60-300 STEM equipped with a Super-X energy dispersive X-ray (EDX) detector and Gatan Enfinum ER electron energy-loss spectroscopy (EELS) spectrometer. TEM was operated at 200 keV with ~30 pA beam current. The convergent semi-angle of the STEM probe was 17 mrad, and the annular dark-field detector inner angle of high-angle annular dark-field (HAADF) images was 55 mrad. For transport characterization, 10 nm thick δ-TaN samples were patterned into Hall bar devices by photolithography and Ar ion milling. The electrical transport of δ-TaN Hall bar devices was tested through DC setup measurement by utilizing a physical property measurement system (PPMS) (Quantum Design, DynaCool).

Devices for ST-FMR measurement were fabricated using microstrips with dimensions 3-40 μm (wide) × 30 μm (long) using contact optical lithography first and then they were etched in an argon ion milling system. We verified that different dimensions here did not affect the measured SHA [57]. The data were obtained mostly using devices with dimensions $15 \times 30$ μm$^2$ employing contact optical lithography. A 3-terminal contact (Ti (7)/Au (150)) was deposited at the two terminals of the bar. The in-plane static magnetic field was generated by a GMW 3D magnet whereas a rf current signal generator was used to generate microwaves with a frequency of 15 to 7 GHz and 2.5 V voltage. A reference sample with the TaN(10) layer replaced by Ta(5) was also prepared. Devices for spin pumping were fabricated also by photolithography with dimensions 600 μm (wide) × 1500 μm (long). The measurement equipment and parameter setup were the same as for the ST-FMR measurements. A reference sample with the structure Ta(5)/CFB(5)/MgO(2)/Ta(2) was prepared to normalize the field value which is generated by the waveguide. The calibration of our waveguide setup can be found in ref. [57].

Electronic structure calculations were performed with the projector augmented wave method for the electron-ion interaction as implemented in the Quantum Espresso code [58,59]. The exchange and correlation effects were treated using generalized gradient approximation (GGA) in the form proposed by Perdew, Burke, and Ernzerhof [60,61]. Spin-orbit coupling (SOC) was included as a second variational step, using scalar-relativistic eigenfunctions as the basis, after the initial calculation converged to self-consistency. The k-point scheme with the 21×21×21 Γ-centered Monkhorst-Pack grids was used in the first Brillouin zone sampling. The electronic band structures were further confirmed by the calculations from Elk code [62,63]. In all the calculations, the experimental lattice parameters were adopted. By projecting the Bloch wave functions to the high-symmetry atomic orbital such as Wannier functions, the tight-binding model Hamiltonian was constructed as implemented in the Wannier90 package [64]. The intrinsic SHCs were then calculated from the model Hamiltonian using the Kubo formula approach in the clean limit [32]:

$$\sigma_{ij}^k = e\hbar \int_{BZ} \frac{d\vec{k}}{(2\pi)^3} \sum_n f_{n\vec{k}} \Omega_{n,ij}^k(\vec{k}), \qquad [1]$$

$$\Omega_{n,ij}^k(\vec{k}) = -2Im \sum_{n' \neq n} \frac{\langle n\vec{k}|J_i^{s,k}|n'\vec{k}\rangle \langle n'\vec{k}|v_j|n'\vec{k}\rangle}{(E_{n\vec{k}} - E_{n'\vec{k}})^2} \qquad [2]$$

where spin Hall conductivity $\sigma_{ij}^k$ is the spin current $j_i^{s,k}$ flowing along the $i^{th}$ direction with the spin polarization along $k$, generated by an electric field ($E_j$) along the $j^{th}$ direction, $j_i^{s,k} = \sigma_{ij}^k E_j$. The SHCs were computed by the integral in the BZ with a 100 × $100 \times 100$ k-grid. The drawing of the crystal structure was produced with the aid of VESTA [65].

### III. RESULTS AND DISCUSSION

**A. XRD: Epitaxy, crystallinity, and lattice constants**

Structural analysis of the $\delta$-TaN (see Fig. 1(a) for the schematic crystal structure) thin films, carried out at room temperature using the traditional $\theta$-$2\theta$ scanning method, revealed diffraction peaks originating from both the MgO (001) substrate and the TaN film. The interface roughness between the MgO (001) substrate and the TaN film is characterized using low-angle XRR. For comparison, the diffractogram as well as the interface roughness for $\delta$-TaN on MgO (111) grown at the same time are also measured. As presented in Fig. 1(b) and 1(c), the TaN-MgO(001) interface possesses a lower roughness (0.901 nm) compared with the TaN-MgO(111) interface (1.339 nm). As seen from Fig. 1(d), both diffractograms are dominated by the respective substrate peaks. Only (002) and (004) TaN peaks for MgO(001) and (111) and (222) TaN peaks for MgO(111) are observed in the diffractograms apart from substrate peaks, indicating textured out-of-plane growth in all cases. The film on MgO(001) (red) shows a higher intensity than MgO(111). Taken together with XRR results, TaN on MgO(001) shows the best structural and topographic characteristics. In accordance with the literature data, our measurement of the crystal structure of $\delta$-TaN from the XRD data yielded a cubic unit cell (space group *Fm*-3*m*, no. 225) [45]. In this structure, both Ta and N occupy the same Wyckoff site, 1*a*, which allows atomic randomness. This feature can be useful for tuning electrical properties by balancing Ta to N ratio and keeping the crystal symmetry fixed. Using the standard Bragg formula, we obtain the lattice parameter value $a = 4.32$ Å. The structural behavior of sputtered samples is consistent with earlier thin films grown by the pulsed laser deposition technique [45].

The rocking curves of a sample were analyzed to estimate the grain size (Fig. 1(b) and (c)) [54]. Instrument broadening was corrected by rocking curves of the single-crystal substrate. The TaN peaks are broad, indicating a small crystallite size. Using the Scherrer equation on the TaN(002) and TaN(111) reflections, the grain size was estimated to be about 10 nm for the samples on MgO(001) and MgO(111), respectively.

**B. STEM characterization of TaN thin films**

The crystalline and compositional structures of the $\delta-$TaN(001) thin films were characterized by employing analytical aberration-corrected STEM equipped with EDX and EELS [67,68]. Cross-sectional HAADF-STEM images of the MgO(001)/TaN(10)/CFB (3.5)/MgO (2.2)/Ta(4) layered stack show uniform thickness of all the layers (Fig. 2(a)). HAADF-STEM image of the δ-TaN layer shows dark-contrast patches along with brighter regions of the lattice contrast (Fig. 2(c)), indicating the presence of amorphous regions in the material. Atomic-resolution HAADF-STEM images of the δ-TaN layers obtained in the [100] and [010] directions showed identical lattice contrast with a square arrangement of atomic columns, demonstrating the cubic structure. EDX elemental maps were also taken to examine the composition of each layer (Fig. 2(b) and 3(a)). The ratio of Ta to N was evaluated to be 1.14±0.15, which is close to composition 44% (N) and 56% (Ta) obtained in the XPS measurement. The elemental line profiles also reveal that compared to the MgO- δ-TaN interface, the δ-TaN /CoFeB interface shows slight interdiffusion of the cations. While a coherent interface between the MgO substrate and the δ-TaN layer is seen, a strong strain field formed due to their 4.8% lattice mismatch is also observed from the strain contrast in a low-angle annular dark-field (LAADF)-STEM image (Fig. 3(c)). EELS core-loss N $K$ edge was obtained from δ-TaN, and the fine structures are compared with N-$2p$ partial density of states (DOS) (Fig. 3(b)). Agreement between experimental and simulated spectra is seen, which indicates both good crystal quality and accuracy of the ab initio calculations in this study (see below).

**C. Electrical resistivity**

The next experimental technique used for the characterization of the δ-TaN(10) thin film on MgO(001) substrate was temperature-dependent resistivity in zero magnetic field, with the results shown in Fig. 4. In the normal state, resistivity $\rho$(T) increases as the temperature is decreased (d$\rho$/dT < 0). Comparing $\rho$(300 K) and $\rho$(10 K), resistivity increases by ~83%. This characteristic was observed earlier in many types of solid states, both nanocrystalline and crystalline, showing normal, super-conducting, and magnetically ordered properties, usually attributed to the disordered scattering of charge carriers as a reduction of the elastic scattering time of conduction electrons. As a consequence, in real materials, atomic randomness can lead to quantum corrections in the resistivity, resulting from stronger electron–electron interaction and weak localization [69-74].

In particular, the temperature dependence of $\rho(T)$ in the range 300 - 50 K demonstrates the weak localization effect (WL), which was previously observed in many thin film materials including superconductors TeSe$_{1-x}$Te$_x$, NbN and TiN [75-77], topological insulators Bi$_2$Se$_3$ and SnTe, topological semimetals Pd$_3$Bi$_2$S$_2$ and WTe$_2$, Kondo systems, and others [78-84]. Therefore, the resistivity behavior can be associated with scattering of charge carriers mainly due to atomic randomness on the Wyckoff occupations between Ta and N positions in the in δ-TaN(10) crystal structure, as well as disorder and lattice defects. However, due to topological features observed in *δ*-TaN along some high symmetry lines in the Brillouin Zone (see below), contributions from electron-electron interactions can also be involved in the scattering process [85,86]. The quasipowerful character of the *ρ*(T) even to the high-temperature data may suggest that there is no activation-type behavior in this system. Based on this, any large gap in the electronic density of states (if it exists) is not expected to be located close to the Fermi energy [87]. The room temperature resistivity value *δ*-TaN is ~ 74 μΩ-cm, which is approximately three times lower than that of *β*-Ta, but comparable to values obtained for *δ*-TaN grown with the pulsed laser deposition technique and pure *α*-Ta [31,44]. For SOT-MRAM applications, low resistivity of the spin-to-charge conversion material is highly desired, since it can protect devices from the current shunting [88]. This effect is frequently observed in SOT thin channels with topological materials and usually leads to an increase in the critical current required for magnetization switching. Therefore, lower resistivity materials combining metallic and non-trivial electronic states are required to induce large spin-to-charge efficiency and reduce the critical switching current density.

At low temperatures, the electrical resistivity drops sharply to zero at $T_c$ = 5.6 K, where $T_c$ is defined as the midpoint of the superconducting transition. The observance of the superconducting phase transition at low temperatures is additional evidence of the high quality of the material studied. The somewhat lower superconducting temperature value obtained in the resistivity measurement compared to the bulk $\delta-$TaN material $T_c$ ~ 8.15 K is likely due to the varying degree of the randomness in Ta and N site occupation in the thin film, thus changing the carrier density rather than $E_F$ [71,74]. A similar effect was observed in the case of isostructural NbN and TiN. In particular, based on earlier studies of TaN, adjusting $N_2$ pressure during the sample growth can potentially lower the resistivity and thus lead to increase of the superconducting temperature [45,90-101].

**D. ST-FMR and spin pumping**

The ST-FMR measurement technique has been used to determine SOT in NM/FM bi-layers with an in-plane magnetic layer [10,102]. As shown in Fig. 5(a), when GHz rf current is injected into the microstrip, spin current will be generated in the $\delta$-TaN via spin Hall effect, which then generates oscillation of the magnetic moment of the CFB layer. The resonance of the magnetic moment and the external field will then generate a DC voltage, which will pass the inductor (Fig. 5(a)) and be measured by the nano voltmeter. The illustration of the multilayer structure of the device is shown in the right part of Fig. 5(a).

Figure 5(b) shows the ST-FMR spectra for $\delta$−TaN(10)/CoFeB(5) under 9GHz rf frequencies, which can be described by:

$$V_{mix} = V_S F_S(H_{ext}) + V_A F_A(H_{ext}) \qquad [3]$$

where $F_S(H_{ext})(F_A(H_{ext}))$ is symmetric (antisymmetric) Lorentzian function with an amplitude $V_S$ ($V_A$). As shown in Fig. 5(b), the symmetric and antisymmetric parts of the Lorentzian function contain information about the damping like torque as well as the field-like torque combining with the external field, respectively. The symmetric and antisymmetric parts of the Lorentzian function can be described by:

$$\frac{\tau_{Oe}+\tau_{FL}}{\tau_{AD}} = \frac{V_{Asym}}{V_{Sym}}\left[1 + (4\pi M_{eff}/H_0)\right]^{-\frac{1}{2}} \qquad [4]$$

where $\tau_{AD}$ ($\tau_{FL}$) is the damping (field) like torque and $\tau_{Oe}$ is the torque generated by the Oersted field. $M_{eff}$ and $H_0$ are the effective magnetization and resonance field, respectively. The $M_{eff}$ is obtained by Kittel formula using: $f = \frac{\gamma}{2\pi}\sqrt{H_0(H_0 + 4\pi M_{eff})}$ (see Fig. 5(c)). This approach allows subtracting the field-like torque contribution and thus obtaining the damping-like SOT efficiency more accurately. Finally, the SHA can thus be calculated by the linear fitting (see Fig. 5(d)) [102]:

$$\frac{\tau_{Oe} + \tau_{FL}}{\tau_{AD}} = \left(\frac{J_S}{J_C}\right)^{-1} \frac{e\mu_0 M_S}{\hbar} t_{CFB} d_{TaN} + \frac{\tau_{FL}}{\tau_{AD}} \quad [5]$$

Using this equation, the damping-like SOT efficiency is around ~0.034. The value of the efficiency is 3 times larger than the earlier reported values of SHA in sputtered NbN, but also comparable to the values reported for many archetypal topological materials and spin Hall metals like Ta, Pd, and Pt [15,103,104]. Note that δ-TaN as a metallic material has a relatively small resistance at RT, thus the current shunting in the FM layer can be excluded. Meanwhile, the smaller resistance also indicates smaller Joule heating. Thus, the SOT switching efficiency is comparable with the materials with relatively large SHA but also large resistivities. The ST-FMR for TaN(20)/CoFeB(3,4,4.5,5,6) samples are also measured, and the resulting damping-like SOT efficiency is ~0.028, which is similar to the TaN(10) samples. The similarities of the damping-like SOT efficiency values of TaN(20) and TaN(10) samples suggests the spin diffusion length of TaN should be smaller than 10 nm [105].

We also measured spin-to-charge conversion by spin pumping in the δ-TaN(10)/CFB (5)/MgO (2)/Ta(2) structure. The samples were patterned into stripes with a waveguide insulated by a 55 nm-thick silicon dioxide layer, as shown in Fig. 6(a). When the frequency of the GHz magnetic field matches with the oscillation frequency of the FM layer under a certain resonance field, the spin current will be generated out of the CoFeB layer and injected into the δ-TaN layer due to the spin pumping effect, where it is then converted to a DC charge current due to the inverse Edelstein and spin Hall effects. The relationship between resonance peak versus the field is shown in Fig. 6(b) and can be divided into a symmetric and an asymmetric Lorentzian function by:

$$V_{total} = \frac{V_S \Delta H^2}{\Delta H^2 + (H_{ext} - H_0)^2} + \frac{V_A(H_{ext} - H_0)}{\Delta H(\Delta H^2 + (H_{ext} - H_0)^2)} \qquad [6]$$

where $\Delta H$ is the line width, $H_0$ is the resonance field, $H_{ext}$ is the applied external magnetic field, and $V_S$ ($V_A$) is the symmetric (antisymmetric) voltage component. The antisymmetric part originates from the anisotropic magnetoresistance (AMR) and anomalous Hall effect (AHE) of the CFB layer, while the symmetric component originates from spin-to-charge conversion. The effective magnetization of the CFB layer is obtained by the Kittel formula, which is shown in Fig. 6(c). With the symmetric resonance voltage, we can thus get the charge current density generated by spin-to-charge conversion:

$$Jc = Vsc/(Rw) \qquad [7]$$

where $R$ is the resistance and $w$ is the width of the stripe. With established methods, we can obtain the damping constant $\alpha = 0.0036$. The spin mixing conductance $g_{\uparrow\downarrow} = 2.4 \times 10^{18} \Omega^{-1} m^{-2}$ is calculated with the damping constant and the intrinsic damping constant of CFB ($\alpha_0 = 0.003$). The spin current density $J_S$ is obtained with the following equation:

$$J_S = \frac{g_{\uparrow\downarrow} \gamma^2 h_{rf}^2 \hbar}{8\pi a^2} \left( \frac{4\pi M_S \gamma + \sqrt{(4\pi M_S \gamma)^2 + 4\omega^2}}{(4\pi M_S \gamma)^2 + 4\omega^2} \right) \frac{2e}{\hbar} \qquad [8]$$

where $h_{rf}$ is the microwave RF magnetic field generated by the waveguide, which is obtained from Ampere's law, ω=$2\pi f$ is the excitation frequency, and $\hbar$ is the reduced Planck constant. Finally, the spin-to-charge conversion ratio is calculated by:

$$\eta = \frac{J_C}{J_S L\, Tanh\left(\frac{t_{TaN}}{2L}\right)} \tag{9}$$

where $L$ is the spin diffusion length, which is smaller than ~10 nm in the TaN layer (we used 8 nm for our calculation), and $t_{TaN}$ is the thickness of the TaN layer. Note that from Eq. 9, a larger spin diffusion length will result in a smaller spin to charge conversion efficiency, thus the 8 nm spin diffusion length estimation will not overestimate the conversion ratio. The so-calculated spin-to-charge conversion by spin pumping ratio is 0.031±0.009, which agrees very well with the value extracted from the ST-FMR. However, note that the spin-to-charge conversion will occur both at the interface and in the bulk for semimetal systems. If we consider the interfacial inverse Edelstein effect (IEE) rather than the spin Hall effect, the IEE length can be as large as 1.3 nm, which is sufficiently large for such material systems.

**E. Electronic Band Structure**

To better understand the connections between electronic properties and effectiveness of spin current in $\delta$-TaN phase, we performed systematic first-principles calculations and compared them with the results of the isostructural $\theta$-TaN phase. First, we intentionally exclude the SOC effect to understand the fundamental properties of TaN phases. The electronic band structure of $\delta$-TaN (left panel on Fig. 7(a)) shows clear metallic behavior with several bands crossing the Fermi level. There exist a few gapless

Dirac nodes at high-symmetry *k*-points around the Fermi level, e.g., *W* and *Γ* points, and also the high degeneracy points along the *K-Γ* path. Those Dirac nodes are usually responsible for the topological behaviors and their related exotic electronic transport properties. As the SOC is included (left panel on Fig. 7(c)), those Dirac points become gapped and induce large Berry curvature near the gap opening points (Fig. 7(d)). As can be gathered from eqs. 1-2, the value of Berry curvature in the *k*-space is inversely proportional to the gap size. By fitting the DFT calculated band structure into an effective tight-binding model using the Wannier90 package, we were able to calculate its SHC and analyze the spin Berry curvature contribution from the band-resolved spin Berry curvature plot. As can be seen from the energy-dependent SHC results (right panel to Fig. 7c), there is a sizable SHC of around -240 $(\hbar/e)S/cm$ near the Fermi level. which is considerably strengthened by about 100 $(\hbar/e)S/cm$ in comparison to pure *α*-Ta. Since the SHA can be expressed as [31]:

$$\theta_{SH} = \frac{e}{\hbar}\frac{\sigma_{xy}^z}{\sigma_{xx}} \quad [10]$$

where $\sigma_{xx}$ is the longitudinal charge conductivity, $\sigma_{xy}^z$ is the transverse SHC, the so-obtained experimentally larger absolute value SHA ( $\theta$ $_{δ\text{-TaN}}$ / $\theta$ $_{Ta}$ ~2.6) may indicate a slightly shifted position of the Fermi level combined with somewhat larger resistivity in our *δ*-TaN. Indeed, around 1.8 eV above the Fermi energy, the SHC reaches its peak value of ~ 450 $(\hbar/e)S/cm$, which is even larger than the value of SHC for the *β*-Ta phase. There are several strategies to adjust the Fermi position and maximize SHC. For example, the Fermi level tuning in heterostructures can be achieved by doping via other elements, defect control (changing the N$_2$ pressure during growth process), epitaxial thin film growth on different substrates, and a recently proposed mechanism based on the cooperative effect of charge density waves and non-symmorphic symmetry [106-108].

According to our theoretical calculations and the structural symmetry analysis (not shown here), the out-of-plane spin component $\sigma_{zy}^z$ can contribute up to ~50 % (-83 to -138 $\hbar/e$) S/cm) of the total value of SHC in the vicinity of $E_F$. This property of δ-TaN phase can be of great interest in magnetic memory applications since it can help to enable external field-free and low power switching of the out-of-plane magnetization [109-111].

To get more insight on how geometrical rearrangement of Ta and N atoms in the unit cell influences Berry curvature, we carried out also a series of calculations for the isostructural allotrope of $\delta$-TaN, namely $\theta$-TaN (space group $P$-62$m$, no. 189) [23]. The electronic band structure is characteristic of a weak Dirac semimetal (Figs. 7(b) and 7(e)). Due to a more pronounced semimetallic character of electronic band dispersion in the vicinity of Fermi level, $\theta$-TaN tends to exhibit lower conductivity than in the $\delta-$TaN phase. The band structure without SOC (Fig. 7(b)) shows a Dirac nodal line along $\Gamma-$A high symmetry $k$-path right at the Fermi level with a small hole pocket at the $K$ point. The nodal line becomes gapped when the SOC effect is included (Fig. 7(e)). It is interesting to note that the SHC shows a plateau near the Fermi level, corresponding to the gap opening window around the $A$ point. This indicates the large spin Berry curvature contribution from those nodal lines, which is further confirmed by our band-resolved spin Berry curvature results (Fig. 7(f)). The calculations suggest that engineering the electronic band structures of both TaN phases by tailoring crystal structure and atomic randomness, can lead to obtaining materials with promising large spin-torque efficiency.

**III. Summary**

We have successfully fabricated high-quality superconducting $\delta$-TaN(10) on the MgO(001) substrate using the magnetron sputtering technique. The surface morphology and crystal structure were investigated $\delta$-TaN forms in a cubic crystal structure type (space group *Fm-3m*, no. 225) with a refined lattice parameter $a$ = 4.32 Å, in agreement with that grown by pulsed laser deposition technique in ref. [45]. The electrical resistivity confirms the bulk superconductivity with $T_c$ around 5.6 K. Thermodynamically, the $\delta$-TaN phase is stable in the range of several percent of the stoichiometry, making it possible to incorporate more nitrogen into the structure and enhance covalent bonding. This can lead to both lowering charge and increasing spin conductivities by changing $E_F$ position without breaking the crystal symmetry.

The experimental value SHA = 0.034 of $\delta$-TaN at room temperature is almost 2.6 times greater than the value of around 0.014, estimated for pure α-Ta. Moreover, if we compare the value of SHA = 0.0037 of pure Ta obtained by the spin absorption method in the lateral spin valve structures, we observe a difference of one order of magnitude [112]. Despite that, simple direct comparison between SHA of pure α-Ta and $\delta$-TaN is not possible, based on the fact that both cubic materials exhibit comparable values of resistivity at RT, the torque efficiency enhancements in $\delta$-TaN has a rather intrinsic nature associated to differences in the electronic band dispersions of both compounds. In comparison, isostructural and classical Bardeen-Cooper-Schrieffer (BCS) superconductor NbN shows comparable resistivity (65 μΩcm at 220 K), but its SHA is almost three times lower than investigated here α-Ta at RT. In this case, incorporating N atoms into pure Nb (SHA = 0.0087 at 10 K) does not change SHA significantly, as it is observed in α-Ta and $\delta$-TaN. Correspondingly, a similar value of SHA = 0.037 was obtained for topological superconductor candidate $\beta$-PdBi$_2$ [15,93,96], while the authors showed that SHA of pure elemental Bi and Pd is below 0.005. Based on our DFT calculations, the electronic structure of the $\delta$-TaN phase shows few gapless Dirac nodes and high degeneracy points along high-symmetry K-points in the BZ, which are not observed in the band structure of δ-TaN. These non-trivial electronic features are mostly responsible for enhanced Berry curvature in $\delta$-TaN near the Fermi level. Inducing phase transitions from $\delta$-TaN (space group Fm-3m, no. 225) to $\theta$-TaN (space group P-62m, no. 189) will further enhance the Berry curvature in the vicinity of the Fermi level, and thus lead to enhancements of spin torques in the material.

Finally, our DFT calculations suggest a reasonably large out-of-plane in the SHC component, which could potentially facilitate field-free switching in a multi-layered SOT structure. Experimental investigation, as well as optimization of the SOT efficiency in $\delta$-TaN, including doping, alloying and changing Ta vs N ratio to adjust the Fermi level position, are underway.

**Acknowledgments** This project is supported by SMART, one of seven centers of nCORE, a Semiconductor Research Corporation program, sponsored by the National Institute of Standards and Technology (NIST). Parts of this work were carried out in the Characterization Facility, University of Minnesota, which receives partial support from the NSF through the MRSEC (Award No. DMR-2011401) and the NNCI (Award No. ECCS-2025124) programs. Portions of this work were conducted in the Minnesota Nano Center, which is supported by the National Science Foundation through the National Nanotechnology Coordinated Infrastructure (NNCI) under Award No. ECCS-2025124. We thank Javier Garcia Barriocanal and his team for X-ray diffraction assistance. We acknowledge the MSI at the University of Minnesota for providing the computational resources. P. W. S. is grateful to Adam, Michal, Piotr, and Dominika Swatek for informative and helpful discussion on spin Hall conductivity and its relation with the symmetry and electronic band structure.

**Author contribution** J.-P.W. initiated and supervised the project. P.W.S. and X.H. prepared all the samples, designed the experiments, and carried out the XRD measurement. Y.F., O.J.B., D.Z., P.S., and P.W.S. patterned the devices and carried out ST-FMR and transport measurements. X.H., D.L., and Z.C. conducted XRR measurement and related data processing. T.J.P. performed XPS measurements. H.Y. conducted STEM measurements with input from K.A.M. The first-principle calculations were performed by W.J., T.L., and partially by P.W.S. All the authors analyzed the data. P.W.S. led the writing of the paper and analyzed the data with a portion of the materials provided by Y.F., W.J., D.L., and H.Y., whereas R.P. and D.K. helped edit the final draft of the paper. All the authors discussed the results and commented on the paper.

**Data availability** The data that support the findings of this study are available from the corresponding authors upon request.

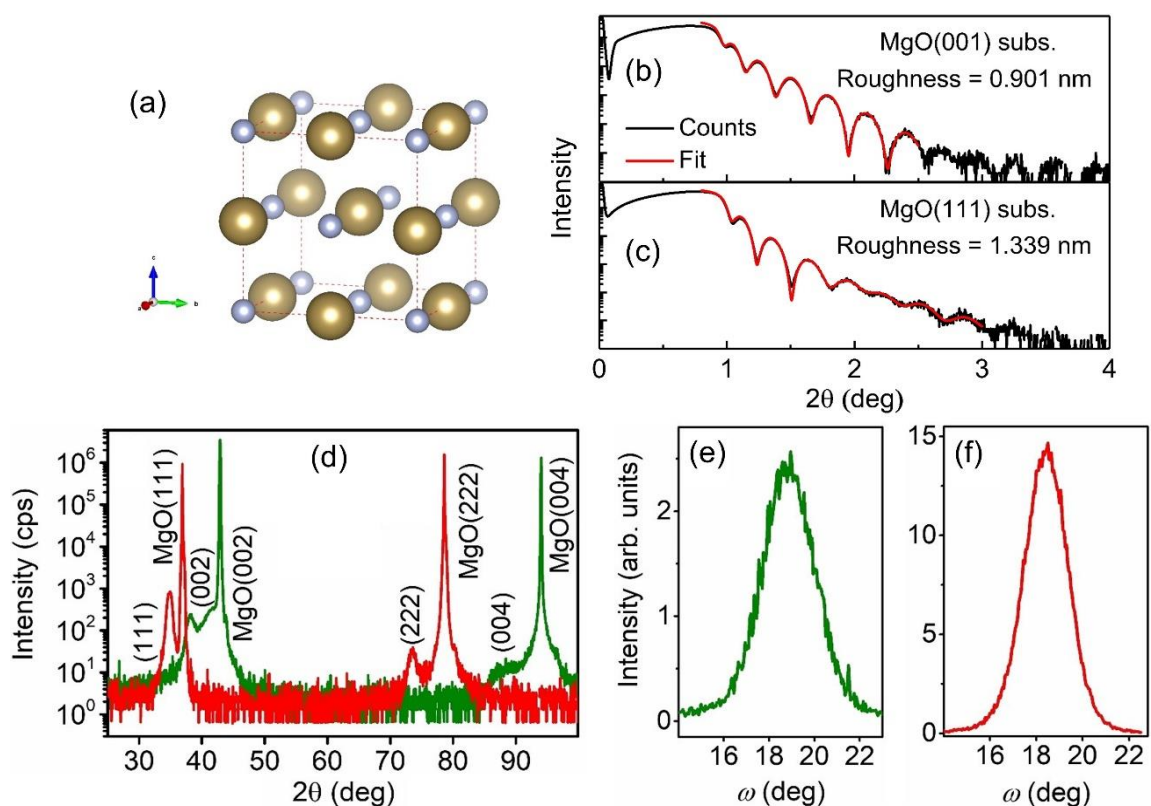

**Fig. 1. Crystal structure, XRR signals with data fitting, and wide-angle XRD patterns of δ-TaN films on MgO (001) and (111) substrates: a)** Schematic crystal structure of *δ*-TaN. Gold balls indicate Ta, purple balls mark nitrogen. **b) and c)** Low-angle XRR data. Fitting results yield a higher roughness at the TaN-MgO(111) interface than that at the TaN-MgO(001) interface. **d)** *θ*-2*θ* scan shows that *δ*-TaN has (001) texture on MgO (001) and (111) texture on MgO (111). **e)** The rocking curve at *δ*-TaN(002) reflection for the sample on MgO(001). **f)** The rocking curve at *δ*-TaN(111) reflection for the sample on MgO(111).

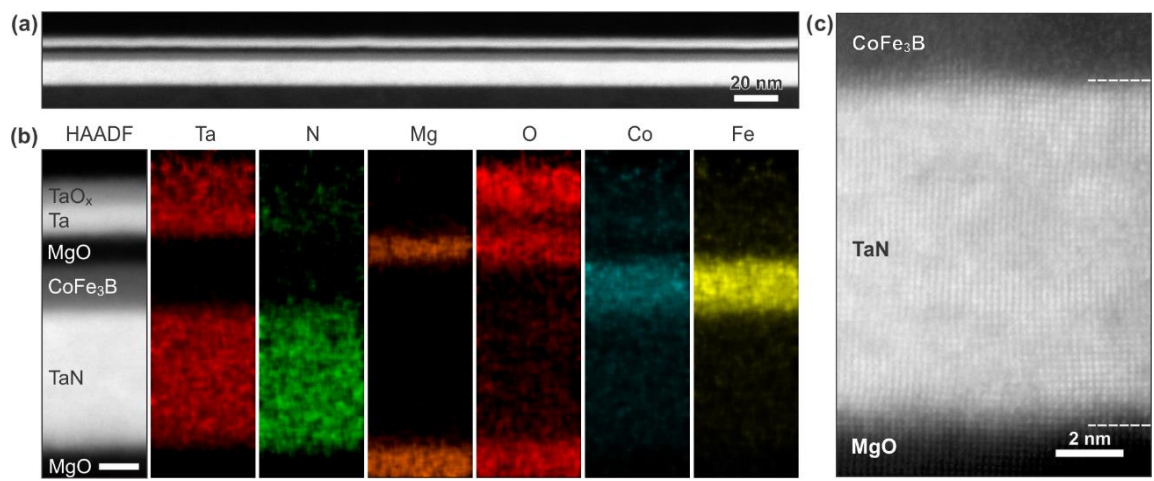

**Fig. 2.** Cross-sectional HAADF-STEM images (**a,c**) and EDX elemental maps (**b**) of a $\delta$-TaN thin film. Scale bar in **b** is 2 nm.

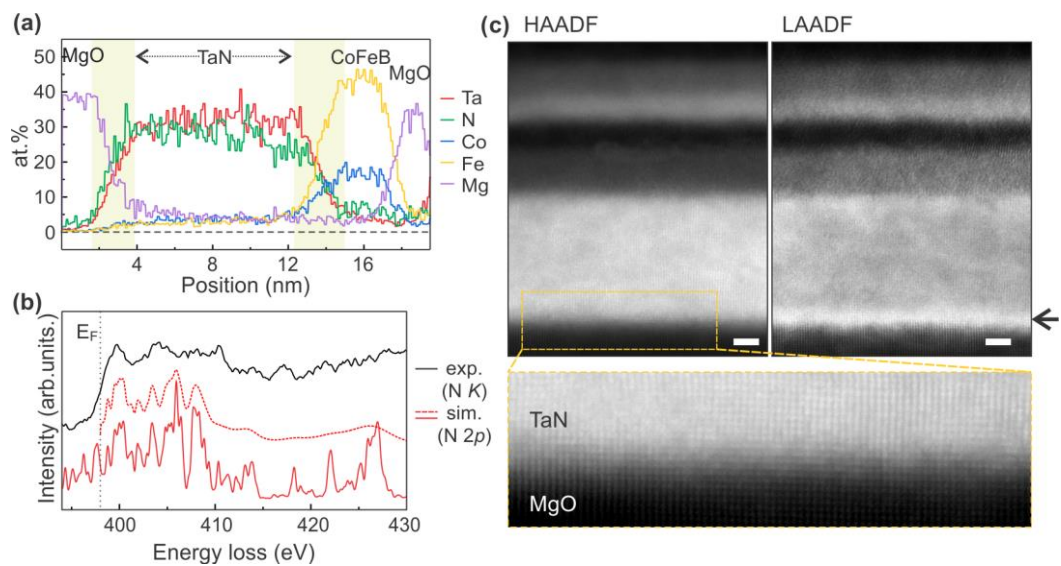

**Fig. 3. a)** EDX elemental line profiles along the film growth direction. MgO-TaN and $\delta$-TaN-CFB interface regions are highlighted with shades. **b)** EELS core-loss N *K* edge acquired from $\delta$-TaN. Fine structures of the edge are compared with N 2*p* partial DOS simulated using *ab initio* calculation. Natural energy broadening was incorporated into the partial DOS and presented as a dashed line. **c)** HAADF- and LAADF-STEM images of a TaN thin film. Strain contrast seen in the LAADF-STEM image is indicated by an arrow. Scale bars are 2 nm.

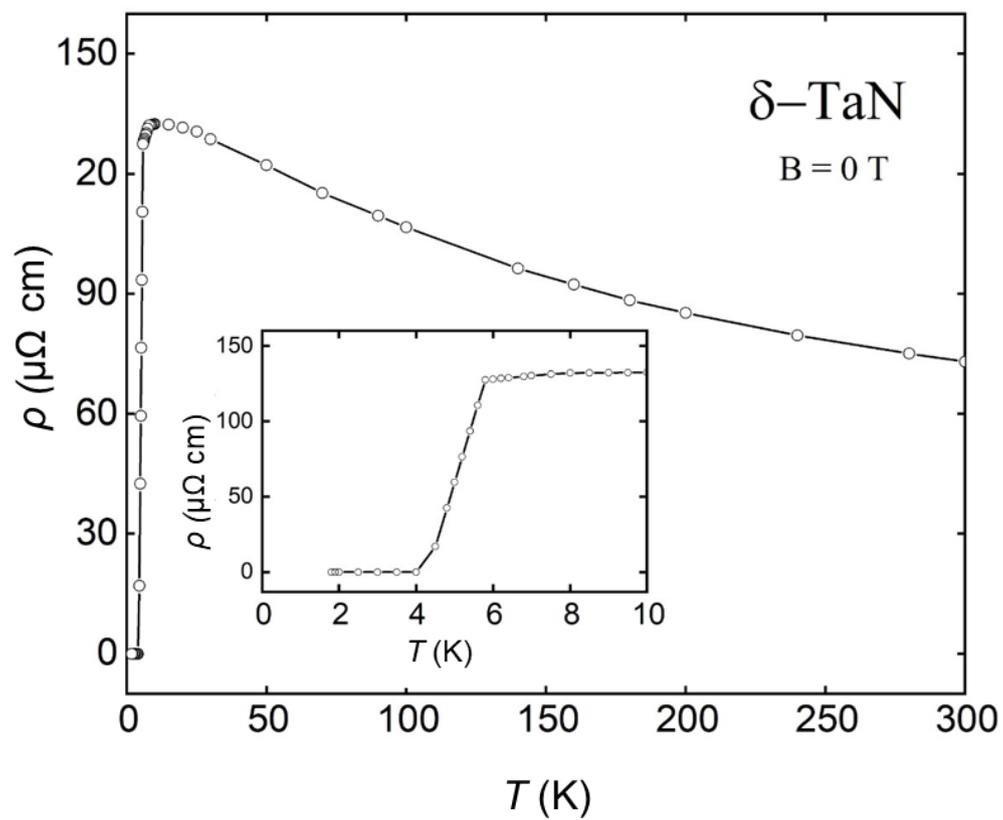

**Fig. 4.** The temperature dependence of the electrical resistivity of $\delta$-TaN measured in the zero magnetic fields. The inset shows the low-temperature resistivity data below 10 K.

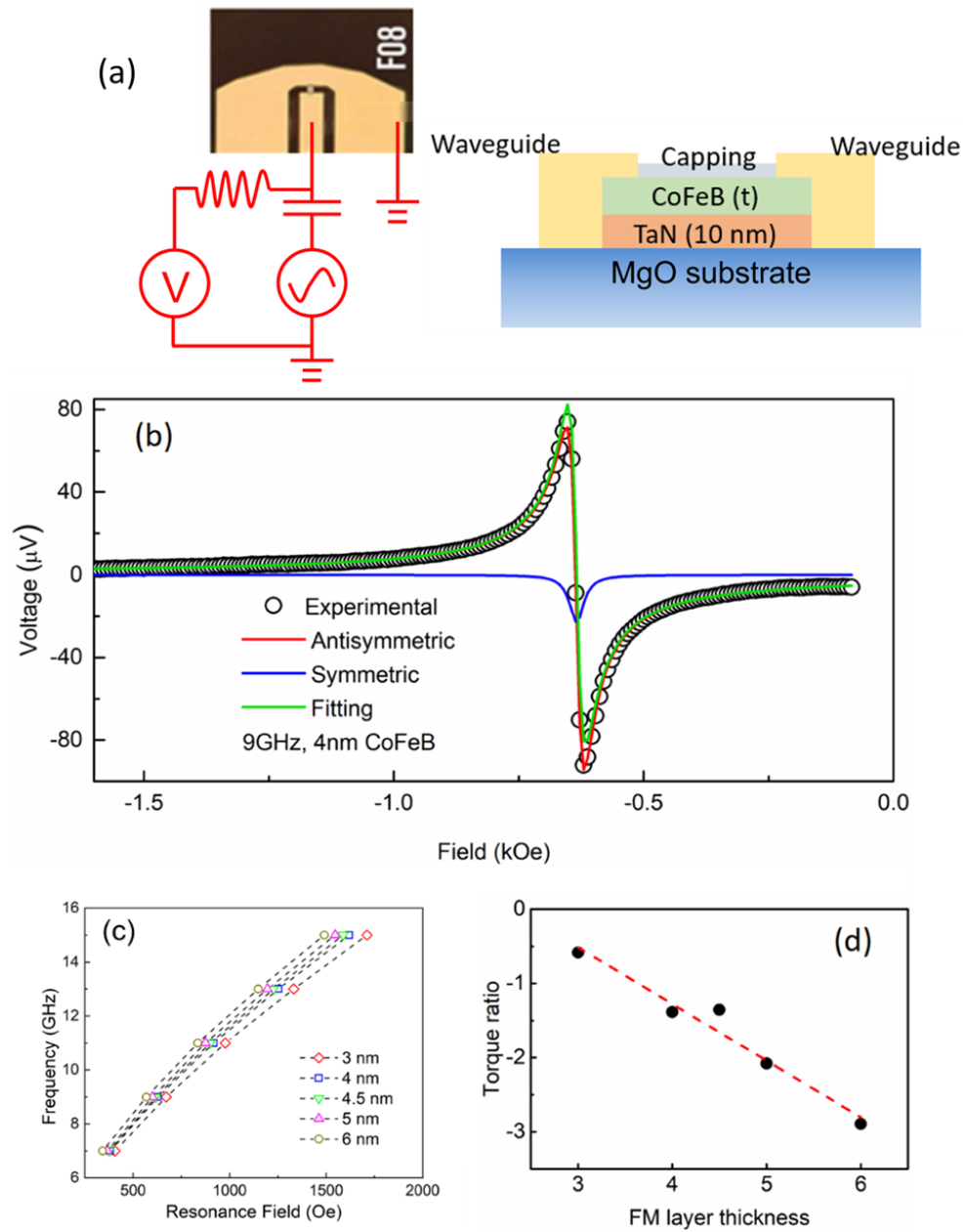

**Fig. 5. ST-FMR measurement of δ-TaN(10)/CoFeB sample: a)** Illustration of device fabrication and measurement setup. The capacitor passes the input rf current and the inductor passes the DC resonance signal. The right figure shows the configuration in yz plane. **b)** The Lorentzian function of ST-FMR measurement, showing the symmetric and antisymmetric parts. **c)** The Kittle fitting of frequency and resonance field for samples with different CoFeB thicknesses. **d)** The linear fitting of $\frac{\tau_{Oe}+\tau_{FL}}{\tau_{AD}}$ versus CoFeB thickness. The spin Hall angle is extracted from this fitting.

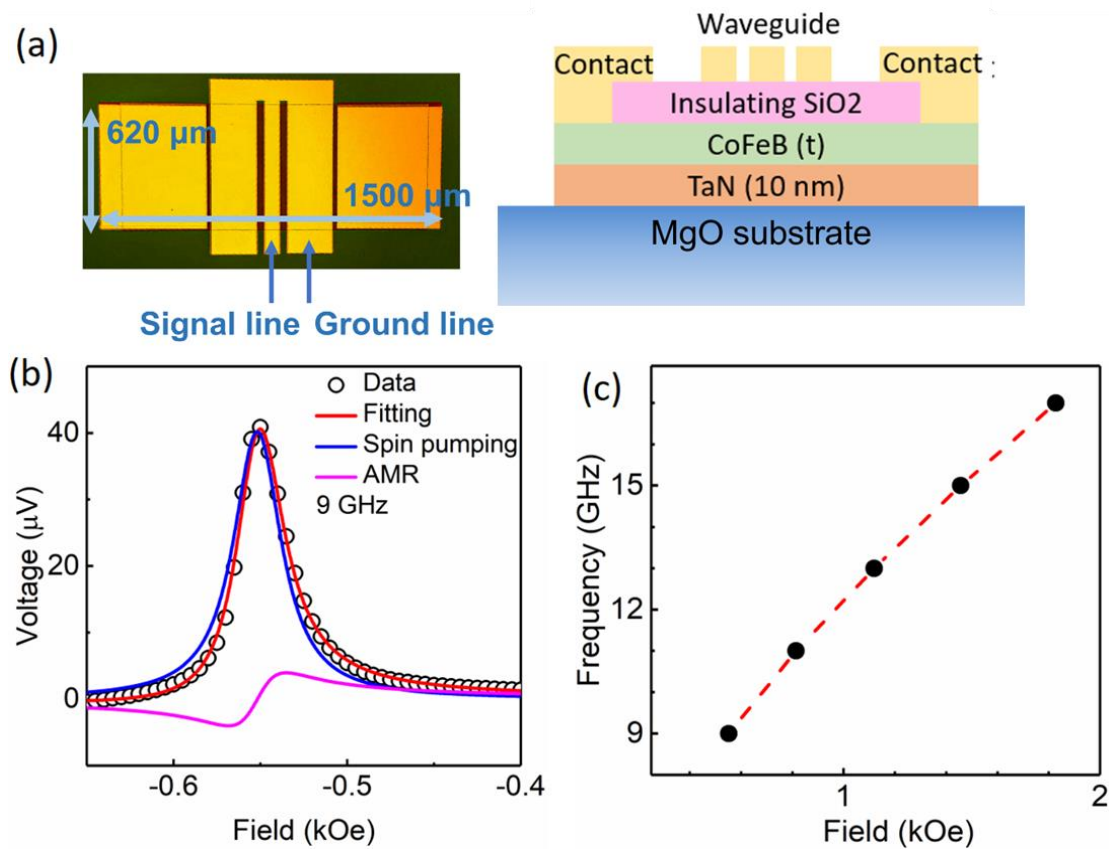

**Fig. 6. Spin pumping measurement of δ-TaN(10)/CFB (5) sample: a)** a) Optical image of the device and illustration of the device structure. The waveguide and the device bar are insulated by 55nm SiO2, which is shown in the right side. **b)** The Lorentzian function of spin pumping measurement, showing the symmetric and antisymmetric parts. Only the symmetric part contains information on spin pumping. **c)** The Kittle fitting of the frequency versus the resonance field.

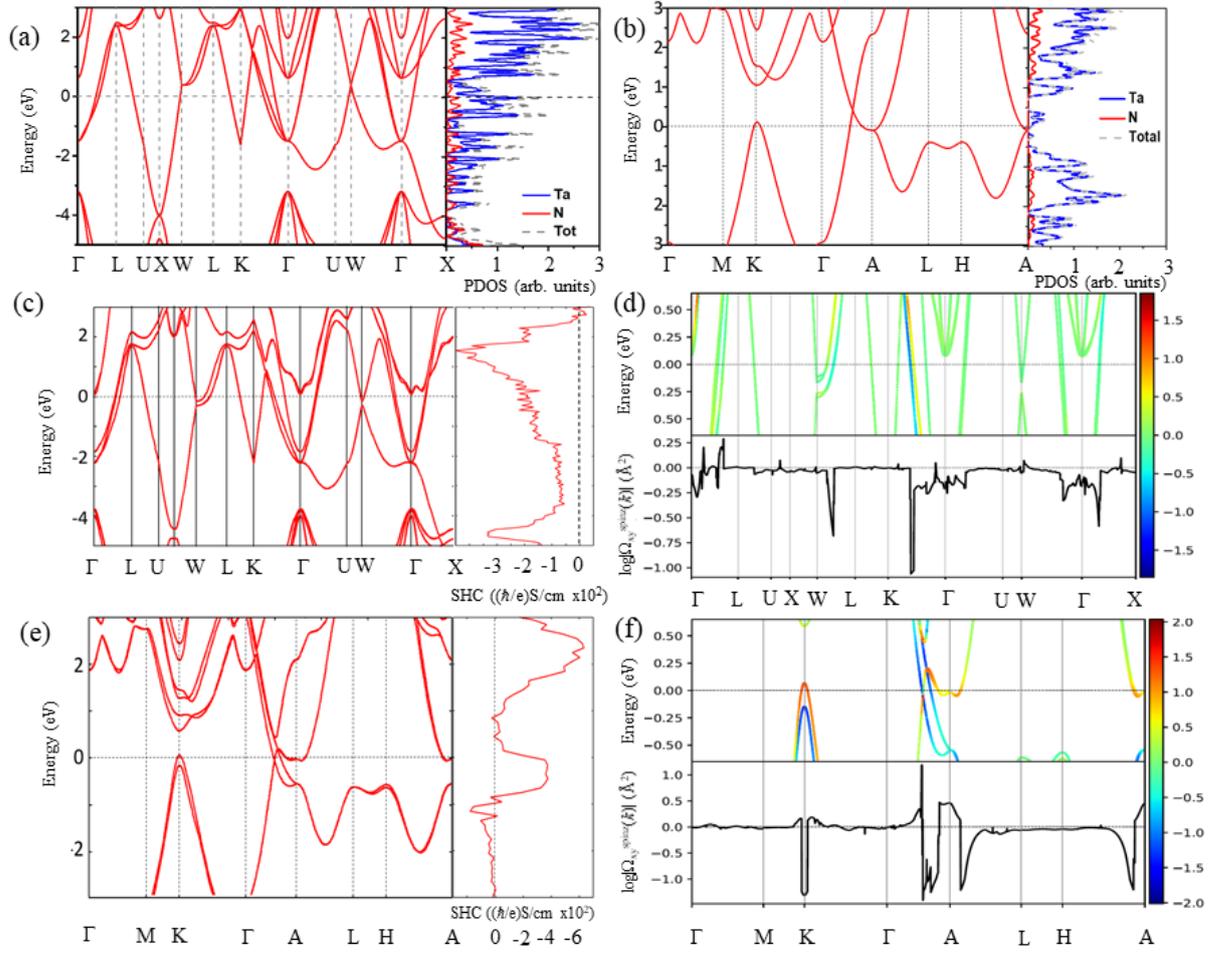

**Fig. 7: First principle calculations: a)** Electronic band dispersion (left panel) and density of states DOS (right panel) without SOC of a) *δ*-TaN and **b)** *θ*-TaN. Electronic band dispersion (left panels) and SHC (right panels) with SOC of **c)** *δ*-TaN and **d)** *θ*-TaN paths in the BZ. **(d)** and **(f)**: The color bar is the SHC projected on each band after taking the logarithm and the *k*-resolved spin Berry curvatures at $E = E_F$.